# STATUS UPDATE ON THE ISAC CONTROL SYSTEM

R. Keitel, D.Bishop, D.Dale, H.Hui, S.Kadantsev, M.Leross, R.Nussbaumer, J.Richards, E.Tikhomolov, G.Waters, TRIUMF, Vancouver, B.C., Canada

## Abstract

Implementation and commissioning of the EPICS based control system for the ISAC radioactive beam facility was completed. The target ion source, mass separator, and low-energy beam-lines started beam production for experiments. In parallel, controls for the accelerator system, a radio-frequency quadrupole followed by a drift-tube linac, and several high-energy beam-lines were implemented and commissioned. The number of controlled devices more than doubled since the last report. An overview of relational database usage, software engineering and quality control methods used will be given.

## 1 ISAC

The ISAC radioactive beam facility at TRIUMF was recently completed and achieved the design specifications. Fully accelerated radioactive beams (1.5 MeV/u) were produced for the first time in October of this year and routine delivery of accelerated beams to experiments has started.

Since the last report [1], the size of the facility increased considerably. In the low energy area, a second electrostatic beam line was added which provides laser-polarized beams. In the high-energy area, the stripper section, a five-tank drift tube linac and three experimental beam lines were added. It should be noted that the ISAC control system also provides controls for vacuum, optics, and diagnostics systems of large experiments, such as the DRAGON spectrometer.

In addition, temporary controls had to be provided for a high current target test with 100 μA of 500 MeV protons.

## 2 CONTROL SYSTEM

The ISAC control system uses the EPICS control system toolkit. Within this framework, a "standard model" architecture is implemented on an Ethernet backbone, using Sun workstation servers, VME based input/output computers (IOCs), and console X-terminals.

### 2.1 I/O Architecture Recap

All ISAC device control is executed through the IOCs. The I/O hardware consists of several subsystems:
- Beam diagnostics devices are controlled via VME modules in order to maintain tight coupling with the IOC CPU [2].
- Beam optics devices have intelligent local controllers, which are supervised by the IOCs via CAN-bus networks [3].
- Vacuum devices are controlled by Modicon Quantum series PLCs, which are peer nodes on the controls Ethernet and are supervised from the IOCs via TCP/IP.
- "Special" devices for laser control use Ethernet based GPIB interfaces [4]
- RF control systems are VXI based. They are peer nodes on the controls Ethernet and are supervised from the IOCs. These systems are maintained by the TRIUMF RF controls group and are outside the scope of this paper [5]

### 2.2 Scope

The recent additions to the ISAC machine more than doubled the number of controlled devices to approximately 2000, and the number of supervised RF systems increased from 1 to 15. The control system contains now 12 IOCs, which interface to 4200 digital, 2500 analog hardware channels, and 40 motors. The EPICS runtime databases grew to a total of 52000 records.

### 2.3 Hardware Additions

Most of the support for the new ISAC sections involved doing "more of the same". In addition, the requirements for laser control and fast faraday cup read-out introduced GPIB-bus devices into the system. They were interfaced with Ethernet based GPIB controllers from National Instruments (GPIP-ENET).

The Ethernet infrastructure was improved by introducing 100BaseT switches in a topology, which allows step-wise isolation of the control system for trouble-shooting and diagnostics. The ISAC controls section of the Ethernet was separated from the rest of the site with a firewall implemented on a Linux system.

## 2.4 Software Additions

Again, a lot of work during the report period involved doing "more of the same", i.e.
- Creating EPICS function block data bases to implement IOC software functionality
- Programming ladder logic for the new vacuum subsystems
- Providing operator interface pages for all new vacuum and optics sub-systems as well the supporting device control panel pages, scripts, save-restore and alarm configurations.

Considerable effort went into automating some of these software activities, in order to improve the quality of the control system and the productivity of the small control system software group. This will be touched upon in the "Quality control" section below.

*Laser stabilization:* The production of Laser-polarized ion beams required integration of GPIB devices and feedback control to stabilize the laser system. This was implemented combining EPICS controls with a LABVIEW system. Although workable, this solution is not robust enough, partially due to the difficulty of integrating LABVIEW into the overall system and some problems with the Ethernet-based GPIB interface. More details about this sub-project can be found in [4].

*Data Archiver:* In trying to archive ISAC data, unhappy experiences both with the EPICS *ar* tool and with a beta version of the high performance binary channel archiver led to the development of a simple archiving tool *trar*, which is tailored to the ISAC requirements:
- ASCII data files, default: one date-stamped file per day
- Archive groups with different disk write intervals
- Conditional archiving of groups
- Possibility of archiving any group to an arbitrary file
- Retaining IOC sampling information by archiving data minimum and maximum between disk write intervals

An accompanying retrieval utility was written, which is wrapped with a Perl-Tk GUI and uses GNUplot for display of the data.

*Macro tool:* A macro tool was developed for capturing operator actions and replaying them. It is implemented as a Perl-Tk script, which uses the output files of the EPICS activity logging facility. The operator presses a "Macro start" button, which inserts a "start recording" message into the activity-logging file. A "Macro stop" button inserts a "stop recording" message. If a "Macro save" button is pressed, the tool extracts all channel access commands between these two messages, retains the ones issued by the macro operator, and formats them into a procedure file, which can be executed by the host based sequencer *cppe* [1].

## 2.5 Operations Support

The ISAC operations console started out with two multi-monitor PC systems under Windows98, which are mainly used as X-terminals using the Exceed X-server from Hummingbird, Inc. Windows98 was chosen because of its multi-monitor capabilities but proved to be unsuitable for routine operations. At least one reboot of the operating system was required per operations shift. Windows98 was replaced with Linux, which increased both performance and reliability dramatically.

The save/restore system was completely re-worked to allow conflict free restoring of beam tunes depending on the selected beam modes for each ion source.

EPICS access control was implemented. Only operations consoles have write access to process variables, but the controls group, beam physicists, and specialists are able to give themselves "self service" write access for a limited time. Experimenters are allowed write access to their own sub-systems and selected devices, dependent on the selected beam modes.

The ISAC controls web site is constantly upgraded with system and trouble-shooting information.

## 2.6 Quality Control

*Relational Data Base:* In the initial stages of the ISAC project, only the interactive EPICS tools were used to generate operator displays (edd display editor) and IOC function-block databases (CAPFAST schematic editor) In order to improve productivity and reduce error frequency, the control system was back-ended with a relational device data base. This database is implemented using Paradox and is supported by several tools implemented as Perl scripts. This system contains all device parameters, interlock specifications and all required information to
- Generate Interlock specification documents for approval by the system specialists
- Automatically generate CAPFAST schematics with device instantiation for ISAC sub-systems [6]
- Automatically generate all device control panels with visualization of device interlocks [7]

- Check the interlock implementation in the PLC programs against the interlock specification [7]
- Generate test documents for sub-system commissioning
- Generate web pages for documentation of module and channel usage in VME crates.

*Bypass/Force summaries:* The control system generates summary information about bypassed device interlocks and forced device states, both overall and by sub-system. The corresponding IOC databases are now automatically generated by Perl scripts, which collect all necessary EPICS record information from the IOC sub-system databases.

*Wiring documentation:* ISAC device wiring diagrams are semi-automatically maintained on the ISAC controls web site. A Perl tool was developed which generates HTML pages using screen dumps of EPICS OPI pages. The OPI call-up buttons on these pages, which would start device control panels on the controls console, are automatically turned into hyperlinks for calling up device wiring diagrams in Adobe PDF format. Buttons, which navigate between OPI pages, maintain this functionality on the web pages.

## ACKNOWLEDGEMENTS

The authors gratefully acknowledge the help of C. Payne from the ISAC operations group in installing and maintaining the group's Linux workstations and firewalls. K. Langton, K. Pelzer, and D. Boehne helped with the hardware installation. Thanks also to the ISAC operations crew and machine physicists for their input and patience while the system was and is being debugged.

## REFERENCES

[1] R. Keitel, et al., 'Design and Commissioning of the ISAC Control System at TRIUMF', ICALEPCS99, Trieste, October 1999, p. 674
[2] D. Bishop et al., 'Custom VME Modules for TRIUMF/ISAC Beam diagnostics', ICALEPCS99, Trieste, October 1999, p.226
[3] D. Bishop et al., 'Distributed Power Supply Control Using CAN-bus', ICALEPCS97, Beijing, November 1997, p. 315
[4] R. Nussbaumer, 'The Laser Stabilization Controls for the ISAC Beam Polarizer', this conference
[5] K. Fong et al., 'RF Control Systems for the TRIUMF ISAC RF', APAC01, to be published
[6] R. Keitel, 'Generating EPICS IOC Data Bases from a Relational Data Base – A Different Approach', this conference
[7] R. Keitel, R. Nussbaumer, 'Automated Checking and Visualization of Interlocks in the ISAC Control System', this conference